\documentclass{PoS}
\usepackage{epsfig}
\usepackage{cite}

\title{Lattice study of the Boer-Mulders transverse momentum distribution
in the pion}

\ShortTitle{Lattice study of the Boer-Mulders transverse momentum distribution
in the pion}

\author{\speaker{M.~Engelhardt}$\,\,^{a\, \dagger} $, B.~Musch$^{b}$,
P.~H\"agler$^{c}$, J.~Negele$^{d}$, and A.~Sch\"afer$^{e}$ \\
        $^{a} $Department of Physics, New Mexico State University, Las Cruces,
        NM 88003, USA\\
        $^{b} $Theory Center, Jefferson National Laboratory, Newport News,
        VA 23606, USA\\
        $^{c} $Institut f\"ur Kernphysik, Johannes Gutenberg-Universit\"at
        Mainz, 55128 Mainz, Germany\\
        $^{d} $Center for Theoretical Physics, Massachusetts Institute of
        Technology, Cambridge, MA 02139, $\mbox{\hspace{0.02cm} } $ USA\\
        $^{e} $Institut f\"ur Theoretische Physik, Universit\"at Regensburg,
        93040 Regensburg, Germany\\
        $^{\dagger} $E-mail: \email{engel@nmsu.edu}}

\abstract{The Boer-Mulders transverse momentum-dependent parton distribution
(TMD) characterizes polarized quark transverse momentum in an unpolarized
hadron. Techniques previously developed for lattice calculations of nucleon
TMDs are applied to the pion. These techniques are based on the evaluation
of matrix elements of quark bilocal operators containing a staple-shaped
Wilson connection. Results for the Boer-Mulders transverse momentum shift
in the pion, obtained at a pion mass of $m_{\pi } = 518\, \mbox{MeV} $,
are presented and compared to corresponding results in the nucleon.}

\FullConference{31st International Symposium on Lattice Field Theory --
LATTICE 2013\\
July 29 -- August 3, 2013\\
Mainz, Germany}

\begin{document}

\section{Introduction}
Transverse momentum-dependent parton distribution functions \cite{revtmd}
(TMDs) encode information about the distribution of transverse (as well as
longitudinal) momentum among partons in a hadron, as extracted from
physical processes such as semi-inclusive deep inelastic scattering
(SIDIS) or the Drell-Yan (DY) process. Isolating hadron structure information
in terms of distribution functions of this type requires a
factorization framework which allows one to disentangle that information
from other components of the process at hand.
For processes containing multiple hadrons in both initial and final states,
factorization breaking contributions may exist \cite{factpos},
the quantitative importance of which is yet unclear. For SIDIS and DY,
on the other hand, factorization does not seem to face serious obstacles,
one possible approach having been advanced in \cite{spl1,collbook,spl2},
although potential issues regarding the analysis of azimuthal
asymmetries have recently been noted \cite{buffmuld}.

\begin{figure}[b]
\vspace{-0.47cm}
\centerline{\psfig{file=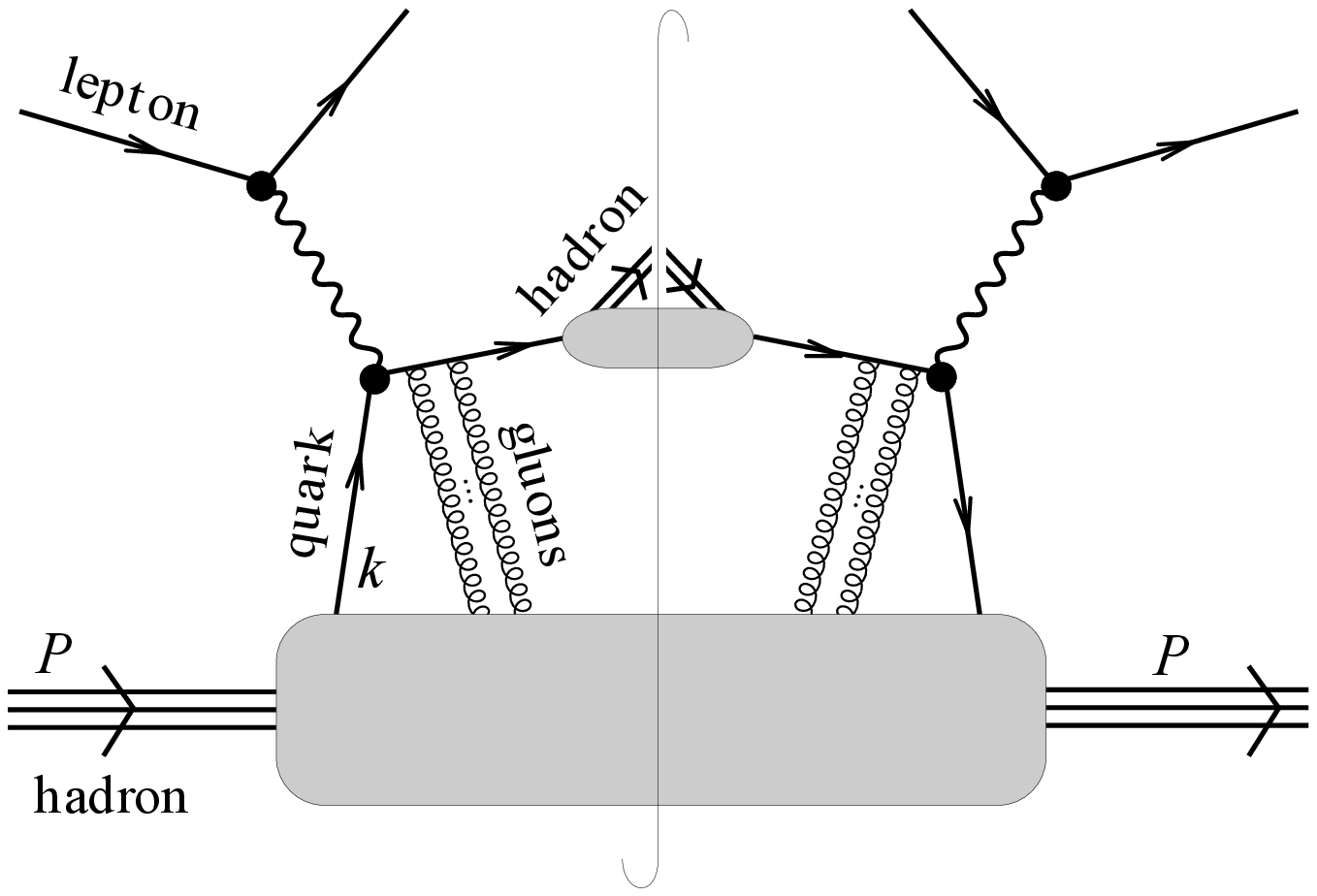,width=7.3cm} }
\vspace{-0.6cm}
\caption{Illustration of the elements of SIDIS factorization. The lower
shaded bubble represents the structure parametrized by TMDs.}
\label{figsidis}
\vspace{-0.37cm}
\end{figure}

A schematic illustration of the principal elements involved in a
description of SIDIS is given by Fig.~\ref{figsidis}; they include
the hard, perturbative vertex, a TMD encoding the structure of the
nucleon, and a fragmentation function describing the hadronization of
the struck quark. It is important to note that factorization does not
necessarily imply that TMDs can be defined completely independent of
the process in which they are embedded. In particular, as also indicated
in Fig.~\ref{figsidis}, final-state gluon exchanges between the struck
quark and the hadron remnant decisively influence the description of
SIDIS (in the DY process, initial-state interactions play an analogous
role). Including these final-state interactions also modifies the
momentum distributions encoded in the TMDs; the manner in which they
formally enter the theoretical definition of TMDs will be elucidated
further below.

These final state effects are important in that they break time-reversal
invariance and thus generate nontrivial T-odd TMDs, leading to corresponding
angular asymmetries in experimental cross sections. Signatures of this kind
have indeed been observed experimentally \cite{hermes}. Nevertheless,
despite the process dependence introduced by accounting for such
effects, a ``modified universality'' across different process types
can be retained in a suitable factorization scheme. Specifically, SIDIS and
DY can be parametrized by the same TMDs, up to a change in sign in
T-odd TMDs \cite{collmetz}.

The project presented here focuses on connecting the phenomenological
definition of TMDs, introduced below, to a concrete lattice QCD
calculational scheme, on the basis of which selected TMD observables
are evaluated from first principles. In particular, exploratory results
for the T-odd Boer-Mulders transverse momentum shift in the pion are given,
thus expanding on the initial study of nucleon TMDs reported in \cite{tmdlat}.

\section{Definition of TMD observables}
The fundamental correlator defining quark TMDs is of the form
\begin{eqnarray}
\Phi^{[\Gamma ]} (x,k_T ,P,S,\ldots ) &=&
\int \frac{d^2 b_T }{(2\pi )^2 } \int \frac{d(b\cdot P)}{(2\pi )P^{+} }
\exp \left( ix(b\cdot P) -ib_T \cdot k_T \right)
\left. \frac{\widetilde{\Phi }^{[\Gamma ]}_{\mbox{\scriptsize unsubtr.} }
(b,P,S,\ldots )}{\widetilde{\cal S} (b^2 ,\ldots )}
\ \right|_{b^{+} =0} \ \ \ \ \ \ \ \ \,
\label{fundcorr}
\end{eqnarray}
with
\begin{equation}
\widetilde{\Phi }^{[\Gamma ]}_{\mbox{\scriptsize unsubtr.} } (b,P,S,\ldots )
\equiv \frac{1}{2} \langle P,S | \ \bar{q} (0) \
\Gamma \ {\cal U} [0,\eta v, \eta v+b,b] \ q(b) \ |P,S\rangle
\label{spacecorr}
\end{equation}
The standard phenomenological description employs a Lorentz frame in
which the hadron of mass $m_h $ propagates with a large momentum in
3-direction, $P^{+} \equiv (P^0 +P^3 )/\sqrt{2} \gg m_h $; then, the
quark momentum components scale such that the correlator (\ref{fundcorr})
and TMDs derived from it are principally functions of the quark
longitudinal momentum fraction $x=k^{+} /P^{+} $ and the quark
transverse momentum vector $k_T $, with the dependence on the component
$k^{-} \equiv (k^0 -k^3 )/\sqrt{2} \ll m_h $ becoming ignorable in this
limit. Correspondingly, (\ref{fundcorr}) is regarded as having been
integrated over $k^{-} $; thus, in the Fourier transform, the conjugate
component $b^{+} $ is set to zero, as written. The hadron momentum and
spin are denoted by $P$ and $S$, and $\Gamma $ stands for an arbitrary
$\gamma $-matrix structure. The ellipsis in
$\Phi^{[\Gamma ]} (x,k_T ,P,S,\ldots )$ indicates that the correlator
will depend on various further parameters, related, e.g., to regularization,
specified below as needed. Heuristically, one can view the
Fourier-transformed bilocal quark bilinear operator as counting quarks
of momentum $k$ in the hadron state, with $\Gamma $ controlling the specific
spinor components involved. However, gauge invariance additionally enforces
the introduction of a gauge connection ${\cal U}$, the precise path of
which will be specified presently. The presence of ${\cal U}$ introduces
divergences additional to the field renormalizations of the quark
operators (this is indicated by the subscript ``unsubtr.''); these
divergences accordingly must be compensated by the additional ``soft
factor'' $\widetilde{\cal S} $. Here, $\widetilde{\cal S} $ will not
need to be specified in detail, since only appropriate ratios in which
the soft factors cancel will ultimately be considered.

The gauge link structure ${\cal U}$ plays a natural role in the correlator
(\ref{spacecorr}), providing a vehicle for incorporating the final state
gluon exchanges between struck quark and hadron remnant discussed in
connection with Fig.~\ref{figsidis}. An effective resummation of these
interactions yields a Wilson line which approximately follows the
trajectory of the struck quark, close to the light cone. The correlator
(\ref{spacecorr}), representing the squared amplitude of the physical
process, thus has parallel Wilson lines attached to both of the quark
operators, extending to large distances along a direction $v$ close to
the light cone; at the far end, these lines are connected by a Wilson line
in the $b$ direction to maintain gauge invariance. The result is the
staple-shaped connection ${\cal U} [0,\eta v, \eta v+b,b]$, where the
path links the positions in the argument of ${\cal U} $ with straight line
segments, and $\eta $ parametrizes the length of the staple. Formally,
it is the introduction of the additional vector $v$ which breaks the
symmetry under time reversal and thus generates T-odd TMDs.

At first sight, the most convenient choice for the staple direction $v$
would seem to be a light-like vector. However, beyond tree level, this
introduces rapidity divergences which require regularization. One
advantageous way to accomplish this is to take $v$ slightly off the light
cone into the space-like region \cite{spl1,collbook}, with perturbative
evolution equations governing the approach to the light cone \cite{spl2}.
This scheme features the modified universality alluded to further above;
the SIDIS and DY processes are connected by inversion of $v$, inducing the
proper sign change in T-odd TMDs. A scheme in which $v$ (along with the
quark operator separation $b$) is generically space-like is also attractive
as a starting point for the development of the lattice QCD calculation, as
will be discussed below. A useful parameter characterizing how
close $v$ is to the light cone is the Collins-Soper evolution parameter
$\hat{\zeta } = v\cdot P / (|v| \, |P|) $, in terms of which the light
cone is approached for $\hat{\zeta } \rightarrow \infty $.

Decomposing the correlator $\Phi^{[\Gamma ]} (x,k_T ,P,S,\ldots )$ into
the relevant Lorentz structures yields the TMDs as coefficient functions.
Whereas in the nucleon case, this leads to eight distinct leading-twist
TMDs, in the simpler $S=0$ pion case investigated here, only two
leading-twist TMDs remain, namely, the unpolarized TMD $f_1 $ and the
T-odd Boer-Mulders TMD $h_{1}^{\perp } $, given, respectively, by
\begin{equation}
\Phi^{[\gamma^{+} ]} = f_1
\ \ \ \ \ \ \ \ \ \ \ \ \ \ \ \ \ \ \ \ \ \ \ \ \ \ \ \ \ \ \ \ 
\Phi^{[i\sigma^{i+} \gamma^{5} ]} =
\frac{\epsilon_{ij} k_j }{m_{\pi } } h_{1}^{\perp }
\label{tmddec}
\end{equation}
The latter characterizes the distribution of transversely polarized
quarks in the (unpolarized) pion.

On the other hand, also the position space correlator
$\widetilde{\Phi }^{[\Gamma ]}_{\mbox{\scriptsize unsubtr.} } $,
cf.~(\ref{spacecorr}), which represents the quantity amenable to
lattice evaluation, can be decomposed analogously in terms of invariant
amplitudes $\widetilde{A}_{iB} $. Again, for $S=0$, only two amplitudes
from the full set obtained for non-zero spin \cite{tmdlat} remain,
\begin{equation}
\frac{1}{2P^{+} }
\widetilde{\Phi }^{[\gamma^{+} ]}_{\mbox{\scriptsize unsubtr.} } =
\widetilde{A}_{2B}
\ \ \ \ \ \ \ \ \ \ \ \ \ \ \ \ \ \ \ \ \ \ \ \ \ \ \
\frac{1}{2P^{+} }
\widetilde{\Phi }^{[i\sigma^{i+} \gamma^{5} ]}_{\mbox{\scriptsize unsubtr.} }
= im_{\pi } \epsilon_{ij} b_j \widetilde{A}_{4B}
\label{ampib}
\vspace{0.06cm}
\end{equation}
These amplitudes are useful in that they can be evaluated in any desired
Lorentz frame, including one particularly suited for the lattice calculation.
In view of (\ref{tmddec}), they are clearly closely related to
Fourier-transformed TMDs. Performing the corresponding algebra, and
specializing, for the purposes of the present investigation, to the
lowest $x$-moment by choosing $b\cdot P=0$, one has
\begin{eqnarray}
\tilde{f}_{1}^{[1](0)} (b_T^2 , \hat{\zeta } ,\ldots , \eta v\cdot P )
& = & 2 \widetilde{A}_{2B} (-b_T^2 ,b\cdot P=0,\hat{\zeta } , \eta v\cdot P)
/ \widetilde{\cal S} (b^2, \ldots ) \label{ampdec1} \\
\tilde{h}_{1}^{\perp [1](1)} (b_T^2 , \hat{\zeta } ,\ldots , \eta v\cdot P )
& = & 2 \widetilde{A}_{4B} (-b_T^2 ,b\cdot P=0,\hat{\zeta } , \eta v\cdot P)
/ \widetilde{\cal S} (b^2, \ldots ) \label{ampdec2}
\end{eqnarray}
where the generic Fourier-transformed TMD is defined as
\begin{equation}
\tilde{f}^{[1](n)} (b_T^2 ,\ldots ) = n! \left( -\frac{2}{m_N^2 }
\partial_{b_T^2 } \right)^{n} \int_{-1}^{1} dx\,
\int d^2 k_T \,
e^{ib_T \cdot k_T } \ f(x,k_T^2 ,\ldots )
\end{equation}
The $b_T \rightarrow 0$ limit formally yields $k_T $-moments of TMDs.
However, this limit contains additional singularities, which one can view
as being regulated by a finite $b_T $. Here, results will only be given
at finite $b_T $. It is important to note the presence of the soft factors
$\widetilde{\cal S} $ on the right-hand sides of (\ref{ampdec1}) and
(\ref{ampdec2}). Absent an evaluation of these soft factors, which
themselves depend on $b^2 $, one cannot directly Fourier-transform the
amplitudes $\widetilde{A}_{iB} $ extracted from a lattice calculation
back to momentum space to obtain the original TMDs defined in (\ref{tmddec}).
On the other hand, one can construct an observable in which the soft factors
cancel by normalizing the (Fourier-transformed) Boer-Mulders function
(\ref{ampdec2}) by the unpolarized TMD (\ref{ampdec1}), which essentially
counts the number of valence quarks. Thus, one defines the ``generalized
Boer-Mulders shift''
\begin{equation}
\langle k_y \rangle_{UT} (b_T^2 , \ldots ) \equiv m_{\pi }
\frac{\tilde{h}_{1}^{\perp [1](1)} (b_T^2 , \ldots )}{\tilde{f}_{1}^{[1](0)}
(b_T^2 , \ldots )} = m_{\pi }
\frac{\widetilde{A}_{4B} (-b_T^2 ,0,\hat{\zeta } ,
\eta v\cdot P)}{\widetilde{A}_{2B} (-b_T^2 ,0,\hat{\zeta } ,
\eta v\cdot P)}
\label{gbmshift}
\end{equation}
which is the regularized, finite-$b_T $ generalization of the
``Boer-Mulders shift''
\begin{equation}
m_{\pi } \frac{\tilde{h}_{1}^{\perp [1](1)}
(0, \ldots )}{\tilde{f}_{1}^{[1](0)} (0, \ldots )} =
\left. \frac{\int dx \int d^2 k_T \, k_y
\Phi^{[\gamma^{+} + s^j i\sigma^{j+} \gamma^{5} ]}
(x,k_T ,P,\ldots )}{\int dx \int d^2 k_T \,
\Phi^{[\gamma^{+} + s^j i\sigma^{j+} \gamma^{5} ]}
(x,k_T ,P,\ldots )} \right|_{s_T =(1,0)}
\label{bmshift}
\end{equation}
which, in view of the right-hand side, formally represents the average
transverse momentum of quarks polarized in the transverse (``$T$'') direction
orthogonal to said momentum in an unpolarized (``$U$'') pion, normalized to
the corresponding number of valence quarks. In the interpretation of
(\ref{bmshift}), it should be noted that the numerator sums over the
contributions from quarks and antiquarks, whereas the denominator contains
the difference between quark and antiquark contributions, thus giving the
number of valence quarks. Note furthermore that ratios of the type
(\ref{gbmshift}) also cancel $\Gamma $-independent multiplicative
field renormalization constants attached to the quark operators in
(\ref{spacecorr}) at finite physical separation $b$.

\section{Lattice evaluation and results}
The formal framework laid out above provides all the necessary elements
for a lattice QCD evaluation of the generalized shift (\ref{gbmshift}).
The path towards this observable proceeds via the calculation of pion
matrix elements of the type (\ref{spacecorr}), yielding the relevant
invariant amplitudes $\widetilde{A}_{iB} $ via (\ref{ampib}). This
requires a setting in which the four-vectors $b$ and $v$ are generically
space-like: The standard scheme for obtaining matrix elements such as
(\ref{spacecorr}) operates with (ratios of) Euclidean space-time
correlators, in which evolution in Euclidean time serves to suppress
pion excited states between, on the one hand, pion source and sink and,
on the other hand, the operator inserted at an intermediate Euclidean time.
In this scheme, only matrix elements of operators defined at a
single Minkowski time are straightforward to evaluate; finite Minkowski
time separations in the operator cannot be directly accomodated on the
Euclidean lattice. Only if all parts of the matrix element under
consideration can be evolved in time to a single instant does rotation
between Euclidean and Minkowski space become trivial. Consequently,
lattice evaluation of the matrix element (\ref{spacecorr}) requires
generically space-like $b$ and $v$, since only then is there no obstacle
to boosting the problem to a Lorentz frame in which $b$ and $v$ are purely
spatial, and calculating
$\widetilde{\Phi }^{[\Gamma ]}_{\mbox{\scriptsize unsubtr.} } $
in that frame. The results extracted for the invariant amplitudes
$\widetilde{A}_{iB} $ are then immediately valid also in the
original frame in which (\ref{spacecorr}) was initially defined,
thus completing the determination of the shift (\ref{gbmshift}).

Since, in a numerical lattice calculation, the staple extent $\eta $
necessarily remains finite, two extrapolations must be
performed from the generated data, namely,
$\eta \rightarrow \infty $, as well as the extrapolation of the
staple direction toward the light cone, $\hat{\zeta } \rightarrow \infty $.
In a previous investigation of nucleon TMDs \cite{tmdlat}, the former
extrapolation was seen to be under control for a range of parameters,
whereas the latter presented a formidable challenge. The
main limitation in this respect is the set of hadron momenta $P$
accessible with sufficient statistical accuracy. One of the main
motivations for the present pion study was to achieve progress with
respect to the large-$\hat{\zeta } $ limit. The pion, by virtue of its
lower mass compared to the nucleon (note that the hadron mass enters
the denominator of $\hat{\zeta } $), allows one to access higher
$\hat{\zeta } $; also, being spinless, it allows one to obtain better
statistics for the TMD matrix element (\ref{spacecorr}) by averaging
over spatial rotations of the operator under consideration. The
calculations presented in the following were performed using a
MILC 2+1-flavor gauge ensemble \cite{milc} on $20^3 \times 64$ lattices
with a spacing of $a=0.12\, \mbox{fm} $, corresponding to pion mass
$m_{\pi } =518\, \mbox{MeV} $. The largest $\hat{\zeta } $ value
reached is $\hat{\zeta } = 2.03$. Disconnected contributions to the
matrix elements (\ref{spacecorr}) were not evaluated.

\begin{figure}
\psfig{file=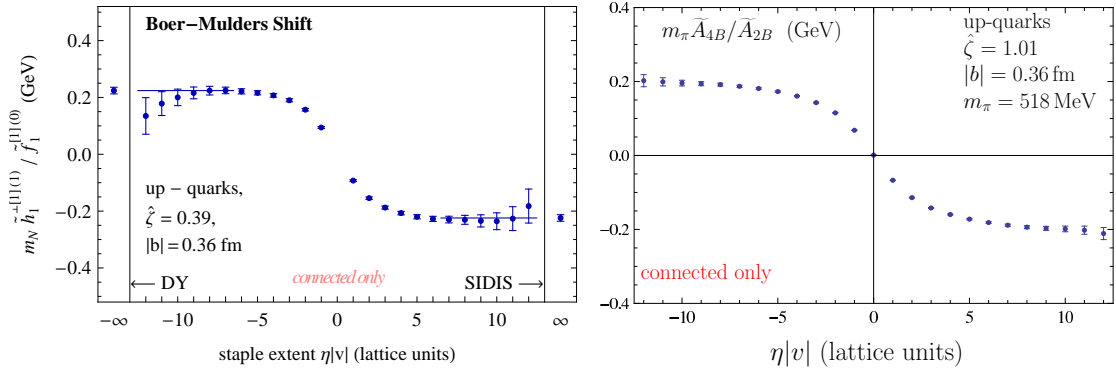,width=15cm}
\caption{Generalized Boer-Mulders shift as a function of staple extent
for $u$-quarks in a proton (left) and a $\pi^{+} $ meson (right). Data
are obtained in the same, $m_{\pi } =518\, \mbox{MeV} $, gauge ensemble
at identical $|b_T |$ and $m_h \hat{\zeta } $.}
\label{compnp}
\end{figure}

Fig.~\ref{compnp} shows a typical result for the generalized Boer-Mulders
shift (\ref{gbmshift}) for $u$-quarks at given $|b_T |$ and $\hat{\zeta } $
as a function of the staple extent, comparing the result for a $\pi^{+} $
meson with the case of a proton studied previously in \cite{tmdlat}.
The T-odd behavior of this observable is evident, with
$\eta \rightarrow \infty $ corresponding to the SIDIS limit, whereas
$\eta \rightarrow -\infty $ yields the DY limit. The data level off to
approach clearly identifiable, stable plateaux as the staple length
grows. Note that the two sets of data correspond to identical hadron
momentum $P$, and the corresponding $\hat{\zeta } $ values differ only
because of the hadron mass in the denominator of $\hat{\zeta } $; i.e.,
$m_h \hat{\zeta } $ is the same in the two cases. In this particular
juxtaposition, the Boer-Mulders shifts are quantitatively very close
to one another, in accordance with a suggestion put forward in
\cite{hannaf}. Note, however, that this observation is special to
$u$-quarks; the $d$-quark Boer-Mulders shift in the proton is
significantly stronger than the $u$-quark shift, whereas the $u$-quark
and $\bar{d}$-quark shifts in the $\pi^{+} $ are, of course, identical.

\begin{figure}
\psfig{file=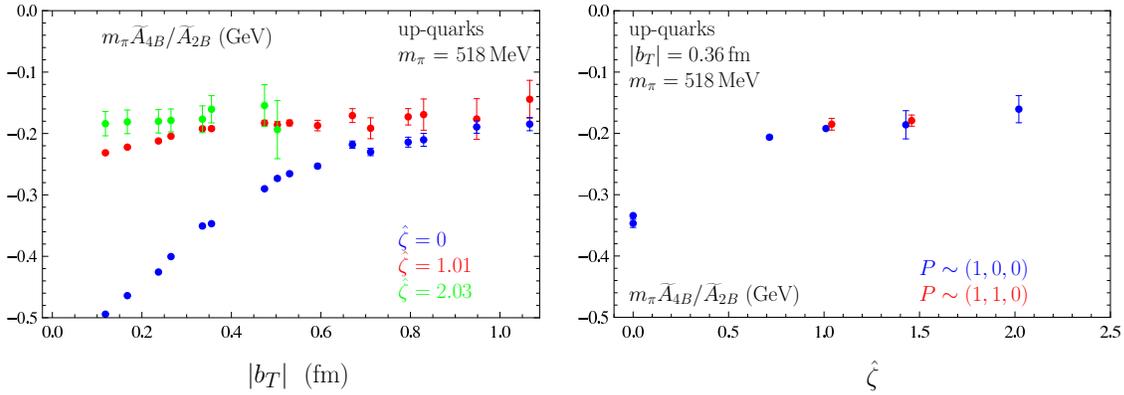,width=15cm}
\caption{Generalized Boer-Mulders shift in the $\eta \rightarrow \infty $
SIDIS limit as a function of $|b_T |$ (left) and $\hat{\zeta } $ (right).
In the left panel, the data in the region below
$|b_T | \approx 0.25\, \mbox{fm} $ may be significantly affected by finite
lattice cutoff effects. In the right panel, the congruence of the data
obtained for $P$ in different directions exhibits the good rotational
properties of the calculation.}
\label{pionbm}
\end{figure}

Fig.~\ref{pionbm} focuses on the data in the SIDIS limit. The left panel
shows the dependence on $|b_T |$ for three different values of the
Collins-Soper evolution parameter $\hat{\zeta } $; the dependence
flattens as $\hat{\zeta } $ is increased. The right panel
exhibits the $\hat{\zeta } $-dependence at a representative $|b_T |$.
A considerably smaller statistical uncertainty is achieved compared to
the nucleon case studied previously \cite{tmdlat}, affording a first
glimpse of asymptotic behavior in $\hat{\zeta } $. The data suggest
a rather early onset of the large-$\hat{\zeta } $ regime, which, if
substantiated further in calculations closer to the physical point,
would be very favorable for the extraction of TMD information
relevant for experiment from lattice calculations.

\section{Summary}
The present study focused on the Boer-Mulders shift in a pion, with a
particular emphasis on obtaining information concerning the behavior of this
type of TMD observable for large Collins-Soper parameter $\hat{\zeta } $.
The Boer-Mulders shift is determined from pion matrix elements of a
quark bilocal operator containing a staple-shaped gauge link which
serves to incorporate final/initial state effects (for SIDIS/DY);
the connection between the Lorentz frame preferred for the lattice
calculation (in which the staple is defined at a single Euclidean
time) and the Lorentz frame preferred for the phenomenological definition
of TMDs (in which the staple direction approaches the light cone from
the space-like side, as parametrized by $\hat{\zeta } $) is achieved by
extracting invariant amplitudes from the data. The results for the
Boer-Mulders shift suggest an early onset of asymptotic behavior as a
function of $\hat{\zeta } $; conclusions about the light-cone limit
thus appear to be within reach of lattice calculations. Furthermore,
the Boer-Mulders shift for $u$-quarks in protons, investigated previously
in \cite{tmdlat}, and $\pi^{+} $ mesons was observed to be quantitatively
similar.

\section*{Acknowledgments}
The lattice calculations performed in this work relied on the Chroma
software suite \cite{chroma} and employed computing resources provided by
the U.S.~DOE through USQCD at Jefferson Lab. Support by the
Heisenberg-Fellowship program of the DFG (P.H.), SFB/TRR-55 (A.S.),
and the U.S.~DOE through grants DE-FG02-96ER40965 (M.E.) and
DE-FG02-94ER40818 (J.N.), as well as through contract DE-AC05-06OR23177,
under which Jefferson Science Associates, LLC, operates Jefferson
Laboratory (B.M.), is acknowledged.

\end{document}